\documentclass[superscriptaddress,showpacs,aps,prd,notitlepage]{revtex4-1}

\usepackage{amsmath,amssymb,color,epsfig}

\begin{document}
\title{Conformastationary disk-haloes  in Einstein-Maxwell gravity}

\author{Antonio C. Guti\'errez-Pi\~{n}eres}
\email[e-mail:]{acgutierrez@correo.nucleares.unam.mx}
\affiliation{Instituto de Ciencias Nucleares, Universidad Nacional Aut\'onoma de M\'exico,
 \\AP 70543,  M\'exico, DF 04510, M\'exico}
\affiliation{Facultad de Ciencias B\'asicas,\\
Universidad Tecnol\'ogica de Bol\'ivar, Cartagena 13001, Colombia}

\begin{abstract}
An exact  solution of  the Einstein-Maxwell field equations for  a  conformastationary  metric  with  magnetized 
disk-haloes   sources is  worked out in full. The  characterization of  the nature of  the  energy momentum tensor of the 
 source  is  discussed.  All  the  expressions  are presented  in terms of  a  solution of  the Laplace's equation. A 
``generalization''  of  the  Kuzmin solution of  the Laplace's equations  is  used as   a  particular  example. The  
solution obtained is asymptotically  flat in general and  turns out to be free of  singularities. All  the  relevant 
quantities show a  reasonable physical behaviour.

\end{abstract}

\maketitle
\section{Introductory remarks}
In  a  recent  work \cite{PhysRevD.87.044010}, we  presented a relativistic model describing a thin 
disk surrounded by a halo in presence of an electromagnetic field. The  model was obtained by solving the Einstein-Maxwell
 equations on a particular conformastatic spacetime background  and by using the distributional approach for the energy-momentum tensor.
The  class  of  solution corresponding to the  model  is asymptotically flat and singularity-free, and satisfies all the energy conditions.
The   purpose  of the present  work is  to  extend   the  above-mentioned  study to the  
conformastationary  case,  and the Kuzmin solution of  the  Laplaces  equation  to include a ``generalized''
Kuzmin  solution of  the Laplace's equation. The reason to  undertake such an endeavour  are easy to  understand. Indeed, 
the issue of  the exact solution of  the Einstein and   Einstein-Maxwell equations   describing isolated sources self  gravitating  
in a stationary axially  symmetric spacetime   appears to be of great interest both from a mathematical and physical point of view.
 For  details of the  astrophysics importance  and  the most relevant developments concerning  the  disks  and disk-haloes sources    
 the  reader is  referred  to the  works  \cite{PhysRevD.87.044010, Chakraborty:2014paa} and references therein. 
 
 In this  work  we  present a new exact solution of the Einstein-Maxwell  field  equations  for a  thin disk surrounded  by a  magnetized 
 halo in  a conformastationary background. This solution is  notoriously simple in its mathematical form. Moreover, the  interpretation
 of  the  energy-momentum tensor presented  here generalises the commonly used  pressure free models to  a fluid with non-vanishing pressure,
 heat flux and anisotropic tensor.   In Section \ref{sec:formalism}  we  present  an  exact general relativistic   model    describing a  disk surrounded
   by an electromagnetized halo  and we obtain  a  solution of  the Einstein-Maxwell field  equations in terms of a solution of Laplace's  equation. 
   In Section \ref{sec:canonicalSEMT}  we   express  the surface  energy-momentum tensor    of the  disk in the  canonical  form and we  present  a physical interpretation
     of   it in  terms  of  a fluid with non-vanishing pressure and  heat flux.  In Section \ref{sec:examples} a particular family of conformastationary magnetized disk-haloes  
   solutions is  presented. We complete  the  paper  with a  discussion of  the  results in Section \ref{sec:conclude}.

\section{General  relativistic magnetized haloes  surrounding thin disks}\label{sec:formalism}
To obtain an exact general relativistic   model    describing a  disk surrounded by an electromagnetized halo in a conformastationary spacetime,   we solved the distributional 
Einstein-Maxwell field equations assuming axial symmetry and that the derivatives of the metric and electromagnetic potential across the disk space-like hyper-surface are 
discontinuous.  To  formulate the corresponding distributional form of the  Einstein-Maxwell field equations,   we   introduce the 
usual  cylindrical coordinates $x^{\alpha} =(t,r,z,\varphi) $ and assume that there exists  an infinitesimally thin  disk located  at  the  hypersurface $z=0$,  so  that the  
metric  and the  electromagnetic  potential can  be written as
      \begin{subequations}
         \begin{eqnarray}
            g_{\alpha\beta} &=& g^+_{\alpha\beta} \theta (z) + g^-_{\alpha\beta} \{ 1 - \theta (z) \},\\
                   &\nonumber\\
             A_{\alpha} &=& A^+_{\alpha} \theta (z) + A^-_{\alpha} \{ 1 - \theta (z) \},
           \end{eqnarray}
      \end{subequations}
respectively. Accordingly, the Ricci tensor reads
              \begin{equation}
                 R_{\alpha\beta} = R^+_{\alpha\beta} \theta(z) + R^-_{\alpha\beta} \{ 1 - \theta (z) \} + H_{\alpha\beta} \delta(z), \label{eq:ricdis}
                \end{equation}
where $\theta(z)$ and $\delta (z)$ are, respectively, the Heaviside and Dirac distributions with support on $z = 0$. Here $g^\pm_{\alpha\beta}$ and $R^\pm_{\alpha\beta}$ are the 
metric tensors and the Ricci tensors of the $z \geq 0$ and $z \leq 0$ regions, respectively, and
      \begin{eqnarray}
         H_{\alpha\beta} = \frac{1}{2} \{ \gamma^z_{\alpha} \delta^z_{\beta}  + \gamma^z_{\beta} \delta^z_{\alpha}
        -\gamma^{\mu}_{\mu} \delta^z_{\alpha} \delta^z_{\beta} - g^{zz} \gamma_{\alpha\beta} \},
      \end{eqnarray}
with  $\gamma_{\alpha\beta} =  2{g_{\alpha\beta,z}}$  and all the quantities are evaluated at $z = 0^+$.  In agreement with  (\ref{eq:ricdis}) the energy-momentum tensor and the 
electric current density are expressed as 
          \begin{subequations}
             \begin{eqnarray}
               T_{\alpha\beta} &=& T^+_{\alpha\beta} \theta(z) + T^-_{\alpha\beta} \{ 1 - \theta(z) \} + Q_{\alpha\beta} \delta(z), \label{eq:emtot}\\
                      & \nonumber\\
                  J_{\alpha} &=& J^+_{\alpha} \theta(z) + J^-_{\alpha} \{ 1 - \theta(z) \} + {\cal I}_{\alpha} \delta(z), \label{eq:eccomp}
                    \end{eqnarray}
                     \end{subequations}
where $T^\pm_{\alpha\beta}$ and $J^\pm_{\alpha}$ are the energy-momentum tensors and electric current density of the $z \geq 0$ and $z \leq 0$ regions, respectively. Moreover, 
$Q_{\alpha\beta}$ and ${\cal I}_{\alpha} $ represent the part of the energy-momentum tensor and the electric current density corresponding to the disk-like source. The 
energy-momentum tensor  $T^\pm_{\alpha\beta}$ in (\ref{eq:emtot}) is taken to be the sum of two distributional components, the purely electromagnetic (trace-free) part and a 
``material'' (trace) part,
            \begin{eqnarray}
                 T^\pm_{\alpha\beta} = E^\pm_{\alpha\beta} + M^\pm_{\alpha\beta},\label{eq:emtotcomp}
              \end{eqnarray}
where $E^\pm_{\alpha\beta}$ is the electromagnetic energy-momentum tensor
            \begin{eqnarray}
               E_{\alpha\beta} = F_{\alpha\nu}F_{\beta}^{\;\,\nu} - \frac{1}{4} g_{\alpha\beta}F_{\mu\nu}F^{\mu\nu}, \label{eq:tab}
               \end{eqnarray}
with $F_{\alpha\beta} =  A_{\beta,\alpha} -  A_{\alpha,\beta}$ and $M^\pm_{\alpha\beta}$  is  an unknown ``material'' energy-momentum tensor  to be obtained. Accordingly, the 
Einstein-Maxwell equations, in geometrized units such that $c = 8\pi G = \mu _{0} = \epsilon _{0} = 1$, are equivalent to the system of equations
     \begin{subequations} \begin{eqnarray}
        G_{\alpha\beta}^{\pm} =R^\pm_{\alpha\beta}  - \frac{1}{2} g_{\alpha\beta} R^\pm &=& E^\pm_{\alpha\beta}
        +  M^\pm_{\alpha\beta}\label{eq:einspm}\\
        H_{\alpha\beta}  - \frac{1}{2} g_{\alpha\beta} H &=& Q_{\alpha\beta}, \label{eq:einsdis}\\
        F^{\alpha\beta}_{ \pm  \ \ ; \beta}   &=&    J^{\alpha}_\pm ,\label{eq:maxext}\\
       \left [F^{\alpha\beta}\right]n_{_{\beta}} &=&   {\cal I}^{\alpha},\label{eq:emcasj}
     \end{eqnarray}\label{eq:EMequations}\end{subequations}
where $H \equiv g^{\alpha\beta} H_{\alpha\beta}$. The square brackets in expressions such as $[ F^{\alpha\beta}]$ denote the jump of $F^{\alpha\beta}$ across of the surface $z=0$  
and  $n_{_{\beta}}$ denotes a unitary vector in the  direction normal  to it. In the  appendix (\ref{sec:einm}) we give the  corresponding  field  equations  and  the 
energy-momentum of the halo  and of the  disk for  a sufficiently general  metric.

To  obtain a  solution of  the  distributional Einstein-Maxwell describing  a system  composed  by a magnetized halo surrounding a  thin disk  in a  conformastationary  
background,  we shall restrict ourselves to  the case where   the  electric potential  $A_t=0$.  We  also  conveniently assume  the  existence  of  a function $\phi$ depending only 
on $r$ and $z$ in  such  a  way that  the metric (\ref{eq:met0}) can be  written in the  form
          \begin{eqnarray}
              ds^2= -e^{2\phi}(dt + \omega d\varphi)^2 + e^{-2\beta\phi} (dr^2 + dz^2 + r^2d\varphi^2) \label{eq:met1},
             \end{eqnarray}
 with $\beta$ an arbitrary  constant.  Accordingly, for  the non-zero components  of  the energy-momentum tensor of  the halo we  have
          \begin{subequations}\begin{eqnarray}
              M_{tt}^{\pm}&=&  -e^{2(1 +\beta)\phi}\big\{ \beta^2 \nabla \phi\cdot\nabla \phi - 2\beta \nabla^2 \phi
                           + \frac{1}{2}r^{-2} e^{2\beta\phi} \nabla A_{\varphi} \cdot \nabla A_{\varphi}
                           - \frac{3}{4}r^{-2}e^{2(1 + \beta)\phi}\nabla \omega\cdot\nabla\omega\big\}\\
              M_{t\varphi}^{\pm}&=& e^{2(1 +\beta)\phi}\Big\{ \frac{\beta}{2}\nabla\omega\cdot\nabla \phi
                                 +  \frac{3}{4}r^{-2}e^{2(1 +\beta)\phi}\omega \nabla\omega\cdot\nabla\omega
                                 - \beta^2\omega \nabla\phi\cdot\nabla\phi
                                 + 2\beta\omega\nabla^2\phi + \frac{3}{2}\nabla\omega\cdot\nabla\phi\nonumber\\
                                &-& \frac{1}{2}r^{-2}e^{2\beta\phi}\omega\nabla A_{\varphi}\cdot\nabla A_{\varphi}
                                 +\frac{1}{2}\nabla^2\omega- r^{-1}\nabla\omega\cdot\nabla r  \Big\}\\
              M_{rr}^{\pm}&=& (1 - \beta)\nabla^2\phi - (1 - \beta)\phi_{,rr} + (\beta^2 - 2\beta)\phi_{,r}^2
                           + \phi_{,z}^2 -\frac{1}{2}r^{-2}e^{2\beta\phi}(A_{\varphi,r}^2- A_{\varphi,z}^2  )\nonumber\\
                          &+& \frac{1}{4}r^{-2}e^{2(1 + \beta)\phi}(\omega_{,r}^2 - \omega_{,z}^2)\\
              M_{rz}^{\pm}&=& \frac{1}{2}r^{-2}e^{2(1 + \beta)\phi}\omega_{,r}\omega_{,z}
                           - (1 - \beta^2 +2\beta)\phi_{,r}\phi_{,z} - (1 -\beta)\phi_{,rz}
                           - r^{-2}e^{2\beta\phi}A_{\varphi,r}A_{\varphi,z},\\
             M_{zz}^{\pm}&=&-\frac{1}{4}r^{-2}e^{2(1 +\beta)\phi}(\omega_{,r}^2 - \omega_{,z}^2) + \phi_{,r}^2
                          - (1 - \beta)\phi_{,zz} + (1 - \beta)\nabla^2\phi  + (\beta^2 - 2\beta)\phi_{,z}^2\nonumber\\
                        & + & \frac{1}{2}r^{-2}e^{2\beta\phi}(A_{\varphi,r}^2 -A_{\varphi,z}^2),\\
             M_{\varphi\varphi}^{\pm}&=&r^2\nabla\phi\cdot\nabla\phi + (1- \beta)r^2\nabla^2\phi
                                      - (1- \beta)r\nabla\phi\cdot\nabla r
                                      - \frac{1}{2}e^{2\beta\phi}\nabla A_{\varphi}\cdot{\nabla}A_{\varphi}\nonumber\\
                                     & + & 
                   e^{2(1 +\beta)\phi}\Big\{\frac{1}{4}(1  + 3 r^{-2}e^{2\beta\phi}\omega^2)\nabla\omega\cdot\nabla\omega
                   - \beta^2\omega^2\nabla \phi\cdot\nabla \phi + 2 \beta\omega^2\nabla^2\phi
                   + \omega\nabla^2\omega \nonumber\\
                  &-& 2r^{-1}\omega\nabla\omega\cdot\nabla r + (3 + \beta)\omega \nabla\omega\cdot\nabla\phi
                   -\frac{1}{2}r^{-2}e^{2\beta\phi}\omega^2\nabla A_{\varphi}\cdot\nabla A_{\varphi}   \Big\}.
                          \end{eqnarray}\end{subequations}
 Whereas    the  non-zero  components  of  the   electric current density  on the  halo has the  form
           \begin{subequations}\begin{eqnarray}
               J^{t}_{\pm}&=& e^{-(1 - 3\beta)\phi} \nabla\cdot \{ \omega r^{-2}e^{ (1 + \beta)\phi} \nabla A_{\varphi}\},\\
                J^{\varphi}_{\pm}&=& e^{-(1 - 3\beta)\phi} \nabla\cdot \{ r^{-2}e^{ (1 + \beta)\phi} \nabla A_{\varphi}  \}.
                  \end{eqnarray}\label{eq:haloCURRENT2}\end{subequations}
The  non-zero components of  the surface  energy-momentum tensor  (SEMT) and  the  non-zero components of  the surface electric
 current density  (SECD) of  the disk  are  given by
           \begin{subequations}\begin{eqnarray}
                S_{t t}   &=& 4\beta e^{(2+ \beta )\phi}\phi_{,z} ,\\
                S_{t \varphi}  &=& e^{(2 +\beta)\phi}( 4\beta\omega\phi_{,z} + \omega_{,z}),\\
                S_{r r}  &=& 2(1 - \beta)e^{-\beta \phi}\phi_{,z},\\
                S_{\varphi\varphi}  &=& e^{(2 +\beta)\phi}\{ (4\beta\omega^2
                + 2(1 -\beta)r^2e^{-2(1 + \beta)\phi})\phi_{,z}  +2\omega\omega_{,z}   \},
                \end{eqnarray}\label{eq:SEMTdisk}\end{subequations}
and
             \begin{subequations}\begin{eqnarray}
                 {\cal J}^{t}&=& r^{-2}e^{3\beta \phi}\omega \left[ A_{\varphi,z}\right],\\
                 {\cal J}^{\varphi}&=& -r^{-2}e^{3\beta \phi} \left[ A_{\varphi,z}\right],
                 \end{eqnarray}\label{eq:diskCURRENT2}\end{subequations}
respectively.  Note  that  all the  quantities  are evaluated on the  surface  of  the  disk. 
We will suppose that there is no electric current in the halo, i.e., we assume that 
         \begin{eqnarray}
                J^{t}_{\pm}=J^{\varphi}_{\pm}=0.
                \end{eqnarray}
Hence   the  system  of  equations (\ref{eq:haloCURRENT2}) is equivalent to  the very  simple system
                \begin{subequations}\begin{eqnarray}
                    \nabla  \omega  \cdot  \nabla A_{\varphi}  &=&0 ,\label{eq:disktcurrent2}\\
                     \nabla \cdot \{  r^{-2} {\cal F}  \nabla A_{\varphi} \} \label{diskvarphicurrent2}&=&0,
                     \end{eqnarray}\label{eq:diskcurrent2}\end{subequations}
where  ${\cal F} \equiv e^{(1 + \beta)\phi}$.  If ${\hat e}_{\varphi}$ is  a unit  vector in azimuthal direction  and
$\lambda$ is  an arbitrary function independent of  the azimuthal coordinate $\varphi$, then one has the  identity
                \begin{eqnarray}
                    \nabla\cdot \{ r^{-1} {\hat e}_{\varphi}  \times \nabla \lambda\} =0. \label{eq:identity}
                     \end{eqnarray}
The  identity (\ref{eq:identity}) may be regarded as the integrability condition for the existence of the function
$\lambda$ defined by
                \begin{eqnarray}
                   r^{-2}{\cal F}\nabla A_{\varphi} = r^{-1}{\hat e}_{\varphi}\times \nabla\lambda.\label{eq:identity2}
                    \end{eqnarray}
Accordingly, the identity (\ref{eq:identity}) implies the  equation
                 \begin{eqnarray}
                  \nabla \cdot \{ {\cal F} ^{-1}\nabla \lambda\}  = 0 \label{eq: auxiliar}
                   \end{eqnarray}
for the  ``auxiliary'' potential $\lambda(r,z)$.  In order  to  have an  explicit  form of the metric function $\phi$
and  magnetic potential $A_{\varphi}$ we suppose that  $\phi$ and   $A_{\varphi}$ depend explicitly on $\lambda$.
Consequently the  equation (\ref{eq: auxiliar}) implies
                \begin{eqnarray}
                          - {\cal F}^{-1}{\cal F}' \nabla \lambda \cdot \nabla\lambda + \nabla^2 \lambda = 0,
                         \end{eqnarray}
where
                \begin{eqnarray}
                          {\cal F}' = (1 + \beta){\cal F}\frac{d\phi}{d\lambda}.
                           \end{eqnarray}
Let  us  assume  the  very  useful simplification
               \begin{eqnarray}
                        {\cal F}^{-1}{\cal F}'=k,
                     \end{eqnarray}
with  $k$ an  arbitrary  constant. Then,  we  have $ {\cal F}=k_3e^{k\lambda}$ and
               \begin{eqnarray}
                        -k \nabla \lambda \cdot \nabla\lambda + \nabla^2 \lambda = 0,
                        \end{eqnarray}
where $k_3$ is  an  arbitrary
constant. Furthermore, if  we  assume the  existence  of  a function
               \begin{eqnarray}
                   U=k_4e^{-k\lambda} + k_5,
                        \end{eqnarray}
with $k_4$ and $k_5$ arbitrary  constants, then
               \begin{eqnarray}
                  \nabla^2U= -kk_4e^{-k\lambda}\{ -k\nabla \lambda\cdot \nabla \lambda + \nabla^2\lambda\}=0.
                   \end{eqnarray}
Accordingly,  $\lambda$  can  be  represented   in terms  of solutions of the Laplace's equation:
               \begin{eqnarray}
                  e^{k\lambda}=\frac{k_4}{U - k_5}, \qquad \nabla^2U=0.
                  \end{eqnarray}
Hence,  the  metric potential $\phi$ can be  written in terms of  $U$ as
                  \begin{eqnarray}
                  e^{(\beta + 1)\phi}=\frac{k_3k_4}{U - k_5} \label{eq:metricpot1}.
                  \end{eqnarray}
To  obtain the  metric  function $\omega$  we  first note  that from (\ref{eq:identity2}) we  have  the  relationship
between  $A_{\varphi}$ and  $\lambda$:
               \begin{eqnarray}
                 \nabla A_{\varphi}= A_{\varphi,r} \hat{e}_{r} + A_{\varphi,z} \hat{e}_{z}
                                  = r {\cal F}^{-1}\hat{e}_{\varphi}\times(\lambda_{,r}\hat{e}_{r}
                                  + \lambda_{,z} \hat{e}_{z}  ).
                           \end{eqnarray}
Then  we  have, $ A_{\varphi,r} =-r{\cal F}^{-1}\lambda_{,r}$ and $ A_{\varphi,z} = r{\cal F}^{-1}\lambda_{,r}$,  or,
in terms  of  $U$
               \begin{subequations}\begin{eqnarray}
                  A_{\varphi,r} &=&k_6rU_{,z} ,\\
                  A_{\varphi,z} &=&-k_6rU_{,r},
                   \end{eqnarray}\label{eq:Avarphi}\end{subequations}
where $k_6=1/(kk_3k_4)$.  Furthermore, with (\ref{eq:Avarphi}) into (\ref{eq:disktcurrent2}) we  arrive  to
               \begin{eqnarray}
                   \omega_{,r}U_{,z} - \omega_{,z}U_{,r}=0,\label{eq:omegagen}
                       \end{eqnarray}
which  admits  the  solution $\omega = k_{\omega}U + k_8$,  with $k_{\omega}$ and $k_8$ arbitrary constants.
As we  know,  the line element (\ref{eq:met1}) must reduce to the Minkowski metric at spatial infinity. This means that the gravitational and magnetic fields vanish at large
 distances from the gravitational source, i.e., it is asymptotically flat. This requires that the  constants $k_3k_4=  - k_5=-1$  and  $k_8=0$.

\section{The SEMT   of  the  disk}\label{sec:canonicalSEMT}
In the  above section we  summarised  the  procedure to  obtain conformastationary axially symmetric relativistic  thin disks   surrounded  by a  material  halo in presence of  a 
magnetic field.  Additionally, we  introduced a  functional relationship dependence between the metric  and  the magnetic  potential and  an harmonic  auxiliary  function in order 
to  obtain a  family of solutions of the distributional Einstein-Maxwell  field  equations.  In short,  we used the inverse method, where a solution of the field equations is taken 
and then the energy-momentum tensor is obtained.  Now, the behaviour of the  energy-momentum tensor obtained  must be investigated  in  order   to   find what  conditions must be  
imposed  over  the solutions and the parameters that appear in the disk-haloes models  in  such a way that the energy-momentum  tensor can  describe  a reasonable physical  
source.  When the  the energy-momentum tensor  is diagonal its interpretation is immediate.  On  the  another  hand,  when the  energy-momentum tensor is non  diagonal its 
physical content can be properly analyzed by writing it in the canonical form. Accordingly, to investigate the  physical content of   the SEMT   of the  disk  we   assume that  it 
 is  possible to  express it in the   canonical  form
        \begin{eqnarray}
               S_{\alpha\beta}&=& (\mu + P)V_{\alpha}V_{\beta} + P g_{\alpha\beta}
                               + {\cal Q}_{\alpha}V_{\beta} + {\cal Q}_{\beta}V_{\alpha}
                               + \Pi_{\alpha\beta}\label{eq:SEMTcanonical},
                 \end{eqnarray}
where  ${\cal Q}_{\alpha}V^{\alpha}={\cal Q}^{\alpha}V_{\alpha}=0$.
Consequently, we can  say that  the  disk is  constituted  by   some mass-energy distribution described by the last surface
energy-momentum  tensor   and $V^{\alpha}$  is the  four velocity of certain observer. Correspondingly, $\mu$, $P$, ${\cal
Q}_{\alpha}$ and $\Pi_{\alpha\beta}$ are then the  energy density, the isotropic  pressure,  the  heat flux and the
anisotropic  tensor on the  surface  of  the  disk, respectively. Thus, it  is  immediate to see  that \cite{herrera2014dissipative}
                                 \begin{subequations}\begin{eqnarray}
                                   \mu &=& S_{\alpha\beta} V^{\alpha}V^{\beta},\label{eq:diskenergy}\\
                                      P&=&\frac{1}{3}  {\cal H}^{\alpha\beta}S_{\alpha\beta},\label{eq:diskpressure} \\
                      {\cal Q}_{\alpha}&=& -\mu V_{\alpha} -S_{\alpha\beta} V^{\beta},\label{eq:diskheatflux}\\
                      \Pi_{\alpha\beta}&=&{\cal H}_{\alpha}^{\;\;\mu}{\cal H}_{\beta}^{\;\;\nu}
                                      ( S_{\mu\nu}  - P{\cal H}_{\mu\nu}),\label{eq:diskanisotropict}
                                          \end{eqnarray}\label{eq:observables}\end{subequations}
where the  projection tensor is defined by 
${\cal H}_{\mu\nu}\equiv g_{\mu\nu} + V_{\mu}V_{\nu}$ and  $\alpha=(t,r,\varphi)$.                                         
 It  is easy to  note that  by choosing the angular velocity  to be  zero in $(\ref{eq:cuadrivelocity})$  we  have then a  fluid 
comoving in our coordinates  system.  Hence,  we  may 
 introduce a  suitable reference  frame in terms  of  the local observers
tetrad (\ref{eq:localobservator})  and (\ref{eq:duallocalobservator}) in the  form
 $\{  V^{\alpha},I^{\alpha},K^{\alpha},Y^{\alpha} \} \equiv     \{h^{\;\;\; \alpha}_{(t)},h^{\;\;\; \alpha}_{(r)},h^{\;\;\; \alpha}_{(z)},h^{\;\;\; \alpha}_{(\varphi)}\}$, 
with  the  corresponding dual tetrad 
$\{ V_{\alpha},I_{\alpha},K_{\alpha},Y_{\alpha} \} \equiv     \{-h_{\;\;\; \alpha}^{(t)},h_{\;\;\; \alpha}^{(r)},h_{\;\;\; \alpha}^{(z)},h_{\;\;\; \alpha}^{(\varphi)}\}$. 
Accordingly, by  using (\ref{eq:diskenergy}), (\ref{eq:diskpressure})
and   (\ref{eq:SEMTdisk}) we  have  for  the  surface  energy density  and the pressure of  the  disk \begin{eqnarray}
                \mu &=&4 \beta e^{\beta \phi} \phi_{,z},
                 \end{eqnarray}
                and
           \begin{eqnarray}
                    P &=& \frac{4}{3}(1-\beta)e^{\beta \phi}\phi_{,z} = \frac{1-\beta}{3\beta} \mu,
             \end{eqnarray}
respectively.  By  introducing (\ref{eq:SEMTdisk}) into  (\ref{eq:diskheatflux}) we obtain  for  the  non-zero components
of the heat flux
                 \begin{eqnarray}
                 {\cal Q}_{\alpha}= - e^{(\beta + 1)\phi}\omega_{,z} \delta_{\alpha}^{\varphi}. \label{eq:heat}
                \end{eqnarray}
Similarly,  by using (\ref{eq:diskanisotropict}) and   (\ref{eq:SEMTdisk}) we  have  for  the non-zero
components  of  the  anisotropic tensor
          \footnote{We  can rewrite the   anisotropic tensor in the form
                    $ \Pi_{\alpha\beta}=P_r K_{\alpha}K_{\beta} + P_{\varphi} Y_{\alpha}Y_{\beta}.$
                    A direct  calculation shows  that
                    $P_r = P_{\varphi} = (1-\beta)\mu/6\beta$}
      \begin{subequations}\begin{eqnarray}
                          \Pi_{rr} &=&\frac{2(1-\beta)}{3}e^{-\beta \phi}\phi_{,z},\\
                          \Pi_{\varphi\varphi} &=&\frac{2(1-\beta)}{3}r^2e^{-\beta \phi}\phi_{,z}=r^2 \Pi_{rr} .
                   \end{eqnarray}\end{subequations}
It is  important  to remark   that  due to that we used the inverse method, no   ``a priori''  restriction are imposed on the physical properties of  the material constituting the 
disks. The  non-zero components of  the SEMT  of  the disks result of ``the nature'' of  the  chosen metric and  the corresponding solutions. So, in our case, the non-zero 
component $S_{rr}$ and  $S_{t\varphi}$  are conditioned by the parameter $\beta$ and  the  metric function $\omega$. When $\beta =1$  the component $S_{rr}$ 
vanishes, whereas $S_{t\varphi}=0$   when $\omega$ vanishes.  The decomposition (\ref{eq:SEMTcanonical})  was  chosen  with the  aim  to describe  the SEMT  by   the  more general  
fluid model.  Hence, the  heat flux appear here in a  ``natural''  way as  a function determined by the metric function $\omega$ and, consequently, by  the  ``rotation''. 
Unfortunately,  as  we  can  see  from (\ref{eq:heat}),  this function  is  oriented along the closed  circular orbits and thus   its physical  interpretation is  unclear. It  is an 
issue which remains unanswered in this manuscript, but  should be addressed in the  future.

Analogously,  the SECD of  the  disk ${\cal J}^\alpha$ can be also  written  in the  canonical  form
          \begin{eqnarray}
                  {\cal J}^{\alpha}&=& { \sigma}V^{\alpha} + {j}Y^{\alpha}  \label{eq:SECDcanonical},
                \end{eqnarray}
then $\sigma$ can be  interpreted as  the  surface electric  charge  density and ${j}$  as  the ``current of magnetization'' of the disk. A direct calculation shows that the   
surface electric  charge  density $\sigma = -V_{\alpha} {\cal J}^{\alpha}=0$,  whereas  the ``current of magnetization'' of the disk  is  given  by
                  \begin{eqnarray}
                   j &=&Y_{\alpha}J^{\alpha}= - r^{-1} e^{2 \beta \phi} \left[  A_{\varphi,z} \right],
                       \end{eqnarray}
where,  as  above, $\left[  A_{\varphi,z} \right]$  denotes  the  jump of  the $z-$derivative  of  the  magnetic potential across  of the  disk and,  all quantities  are  evaluated 
 on the disk. Thus, by  using the results  of  the precedent section, we can write  the  surface energy density, the pressure, the heat flux,  the  non-zero components  of  the 
anisotropic  tensor and the  current of magnetization on the  surface of  the  disk,  respectively, as
             \begin{subequations}\begin{eqnarray}
               \mu &=&  \frac{ 4\beta U_{,z} }
                              { (\beta + 1) (1-U)^{\frac{2\beta + 1}{\beta +1}} }\label{eq:diskenergy+U}\\
                 P&=&\frac{(1 -\beta)}{3 \beta}\mu\label{eq:diskpressure+U},\\
   {\cal Q}_{\alpha}&=& -\frac{k_{\omega} U_{,z}}{1-U} \delta_{\alpha}^{\;\varphi},\label{eq:diskheatflux+U}\\
            \Pi_{rr}&=&\frac{2(1 -\beta)U_{,z}}{3(1+ \beta)(1-U)^{\frac{1}{1+\beta}}}\label{eq:diskanisotropictrr+U},\\
\Pi_{\varphi\varphi}&=& r^2 \Pi_{rr} \label{eq:diskanisotropictvv+U},\\
                             j&=&-\frac{\left[ U_{,r}\right]  }{k (1-U)^{\frac{3\beta}{1+\beta}}} \label{eq:diskcurrent+U},
                                          \end{eqnarray}\label{eq:observables+U}\end{subequations}
where, as  we  know, $U(r,z)$ is  an arbitrary suitable  solution of  the 2-dimensional Laplace's equation in cylindrical  coordinates  and $\left[ U_{,r}\right] $ denotes  of  the 
jump of  the $r$-derivative  of the $U$ across of the  disk. All the quantities are evaluated on the  surface  of  the  disk. It is  important  to  note  that  $k_{\omega}$ is a  
defining  constant in  (\ref{eq:diskheatflux+U}).  Indeed,  when $k_{\omega}=0$  the  heat flux $ {\cal Q}_{\alpha}$ vanishes, a feature of  the  static  disk.

\section{A particular family of conformastationary magnetized disk-haloes  solutions}\label{sec:examples}
In  precedent works \cite{PhysRevD.87.044010, gutierrez2013variational} we have presented  a  model for a conformastatic relativistic thin disk surrounded by a material 
electromagnetized halo from the  Kuzmin  solution  of  the  Laplace's equation in the  form
      \begin{eqnarray}
           U_{K}=-\frac{m}{\sqrt{r^2 + (|z| + a)^2}},\qquad (a, m > 0).\label{eq:ukuzmin}
            \end{eqnarray}
As it is  well known, $\nabla^2 U_K$ must vanish everywhere except on the plane $z = 0$. At points with $z < 0$, $U_K$ is identical to the potential of a point mass $m$ located at 
the point $(r, z) = (0, -a)$, and when $z > 0$, $U_K$ coincides with the potential generated by a point mass at $(0, a)$. Accordingly, it is  clear  that  $U_K$ is generated by the 
surface density of a Newtonian mass
         \begin{eqnarray}
            \rho_K(r,z=0)=\frac{am}{2\pi(r^2 + a^2)^{3/2}}.
            \end{eqnarray}
In this  work,  we  present  a  sort  of  generalisation of  the Kuzmin solution by  considering a  solution  of  the Laplace's  equation in the  form \cite{stephani2003exact},
              \begin{eqnarray}
                    U= - \sum_{n=0}^{N}{\frac{b_n P_n(z/R)}{R^{n + 1}}} ,
                 \qquad P_n(z/R)= (-1)^n\frac{R^{n+1}}{n!}\frac{\partial^n}{\partial z^n}\left(\frac{1}{R}\right)
                    \label{eq:GeneralizedKuzmin},
                  \end{eqnarray}
 $P_n=P_n(z/R)$  being the  Legendre polynomials  in cylindrical  coordinates that was derived in the present form by a direct comparison of    the  Legendre polynomial 
expansion of   the  generating function with  a  Taylor of $1/R$ \cite{arfken1966mathematical}. $R^2\equiv r^2 + z^2$ and  $b_n$  arbitrary constant  coefficients. The 
corresponding magnetic  potential,  obtained  from (\ref{eq:Avarphi}), is
            \begin{eqnarray}
                  A_{\varphi} =- \frac{1}{k} \sum_{n=0}^{N} b_n \frac{(-1)^n}{n!}\label{eq:magpot}
                   \frac{\partial ^n}{\partial z^n}\left(\frac{z}{R}\right)
                  \end{eqnarray}
where,  we  have imposed  $A_{\varphi}(0,z)=0$ in order  to  preserve the regularity  on the axis of  symmetry. Next,  to introduce  the corresponding  discontinuity  in the  
first-order derivatives of  the metric potential and  the magnetic  potential required  to  define the disk we  perform  the  transformation $z \rightarrow |z| +  a$. Thus,  
taking 
 account  of (\ref{eq:observables+U}),  the  surface energy  density of  the  disk,  the  heat flux and  the  non-zero components  of  the  anisotropic tensor  are
        \begin{subequations}\begin{eqnarray}
             \mu(r) &=& \frac{4 \beta \sum_{n=0}^{N}   b_n(n+1){P_{n+1}({a}/{R_a})}{R_a^{-(n+2)}} }
           {(1 + \beta) \left(1 +\sum_{n=0}^N  b_n  P_{n}({a}/{R_a}) R_a^{-(n+1)}\right)^{(2\beta +1)/(\beta + 1)}},\\
         Q_{\alpha} &=& \frac{k_{\omega}\delta_{\alpha}^{\varphi}      \sum_{n=0}^{N}{{b_n P_n(a/R_a)}{R_a^{-(n + 1)}}}}
                   {1  +   \sum_{n=0}^{N}{{b_n P_n(a/R_a)}{R_a^{-(n + 1)}}} },\label{eq:heatkuzgen}
              \end{eqnarray}\end{subequations}
    and
            \begin{eqnarray}
            \Pi_{rr} = \frac{2(1 - \beta)       \sum_{n=0}^{N}  b_n(n+1){P_{n+1}({a}/{R_a})}{R_a^{-(n+2)}}   }
                                       {3(1 + \beta) \left(  1 + \sum_{n=0}^{N}{{b_n P_n(a/R_a)}{R_a^{-(n + 1)}}}
                                       \right)^{1/(1 + \beta)} },\label{eq:stresskuzgen}
                 \end{eqnarray}
respectively. In the  above expressions	 $R_a^2\equiv r^2 + a^2$. As  we know, the other  quantities  are  $P=(1-\beta)\mu/(3\beta)$  and  $ \Pi_{\varphi\varphi} =r^2
\Pi_{rr} $.

The  current of magnetization is
            \begin{eqnarray}
                 j= -  \frac{2r \sum_{n=0}^{N} b_n P_{n+1}'(a/R_a)  R^{-(n+3)}_a}
                      {k (1 + \sum_{n=0}^{N} b_n P_n (a/R_a) R^{-(n+1)}_a )^{2\beta/(1+\beta)}},
                \end{eqnarray}
where  we  have used  (\ref{eq:diskcurrent+U}) and  we first assumed that the  $z$-derivative  of  the  magnetic
potential  present  a finite  discontinuity through  the disk.  In fact, as  we  have  said  above,  the derivatives  of
 $U$ and  $A_{\varphi }$  are  continuous functions across  of  the  surface  of  the  disk.    We  artificially
introduce the  discontinuity through  the  transformation     $z \rightarrow |z| +  a$ .

In order  to illustrate  the  last solution we  consider   particular solutions with $N=0$ and $N=1$. Then we  have
 $U_N$ for the  two first members of  the  family of  the  solutions as  follows,
            \begin{subequations}\begin{eqnarray}
          U_0 &=& - \frac{\tilde{b}_{_0}}{\sqrt{   \tilde{r}^2 +  (|\tilde{z}| + 1)^2}},\label{eq:GeneralizedKuzmin0}\\
          U_1 &=& - \frac{\tilde{b}_{_0}}{\sqrt{   \tilde{r}^2 +  (|\tilde{z}| + 1)^2}}
              \left\{1 +\frac{\tilde{b}_{_1} (|\tilde{z}| +1)}{\tilde{b}_{_0}\left((|\tilde{z}| + 1)^2 + \tilde{r}^2 \right)} \right\}\label{eq:GeneralizedKuzmin1},
            \end{eqnarray}\end{subequations}
 where $\tilde{b}_{_0} = {b}_{_0}/a$ and $\tilde{b}_{_1} = {b}_{_1}/a^2$ whereas $ \tilde{r}=r/a$ and  $\tilde{z}=z/a$.
For the  corresponding magnetic potentials  we  have then
       \begin{subequations}\begin{eqnarray}
          \tilde{A}_{\varphi 0} &=&-\frac{\tilde{b}_{_0}(|\tilde{z}|+1)}{k \sqrt{ \tilde{r}^2 +(|\tilde{z}| + 1)^2}},\label{eq:magpot0}\\
          \tilde{A}_{\varphi 1} &=& -\frac{\tilde{b}_{_0} (|\tilde{z}| + 1)}{k \sqrt{ \tilde{r}^2 + (|\tilde{z}| + 1)^2}}
          \left\{ 1 - \frac{\tilde{b}_{_1}\tilde{r}^2 }
          {\tilde{b}_{_0} (|\tilde{z}| + 1) \left((|\tilde{z}| + 1)^2 + \tilde{r}^2 \right)}    \right\}\label{eq:magpot1},
               \end{eqnarray}\end{subequations}
where $\tilde{A}_{\varphi}=   A_{\varphi}/a$.

 \begin{figure}
$$\begin{array}{cc}
\epsfig{width=3.0in,file=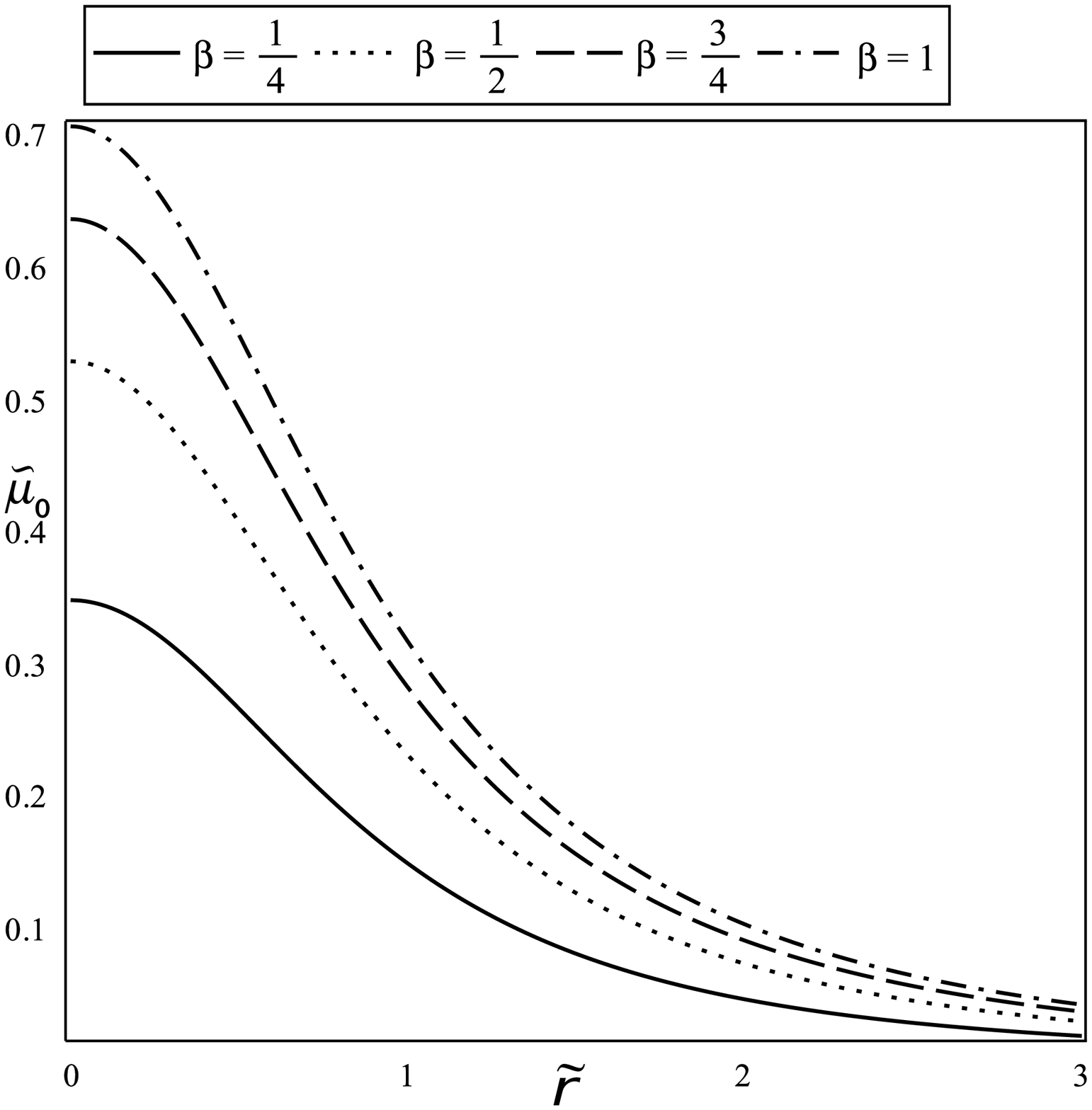}&
\epsfig{width=3.09in,file=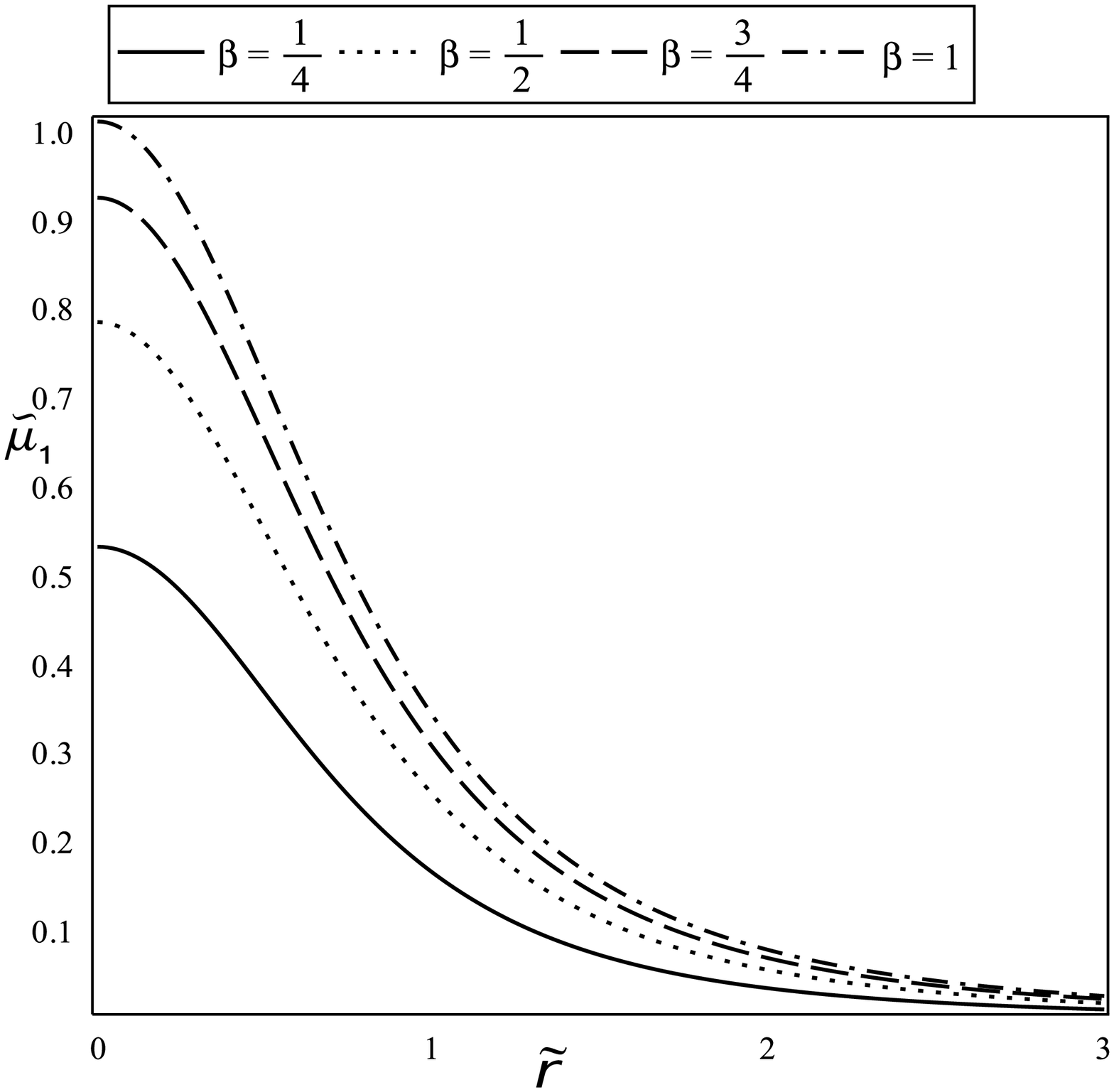}\\
(a) & (b)\\
\end{array}$$
\caption{\label{fig:figure1} Dimensionless surface energy densities  ${\tilde \mu}$  as a function of ${\tilde r}$.
In each case, we plot ${\tilde \mu}_0(\tilde r)$  and ${\tilde \mu}_1(\tilde r)$ for different values of
the parameter $\beta$  with $\tilde{b}_0=1$ and $\tilde{b}_1=0.5.$}
\end{figure}

In Fig. \ref{fig:figure1}, we show the dimensionless surface energy  densities  ${\tilde \mu}$ as a function of ${\tilde
r}$. In each case, we plot ${\tilde \mu}_0(\tilde r)$ [Fig. \ref{fig:figure1}(a)]  and ${\tilde \mu}_1(\tilde r)$ [Fig. \ref{fig:figure1}(b)] for different values of the parameter 
$\beta$  with $\tilde{b}_0=1$ and $\tilde{b}_1=0.5$.
It can be seen that the surface energy density is everywhere positive fulfilling the energy conditions. It can be observed that for all the values of
${\beta}$ the maximum of the surface energy density occurs at the center of the disk and that
it vanishes sufficiently fast as $\tilde r$ increases. It can also be observed that the surface energy
density in the central region of the disk increases as the values of the parameter $\beta$ increase.
 \begin{figure}
$$\begin{array}{cc}
\epsfig{width=3in,file=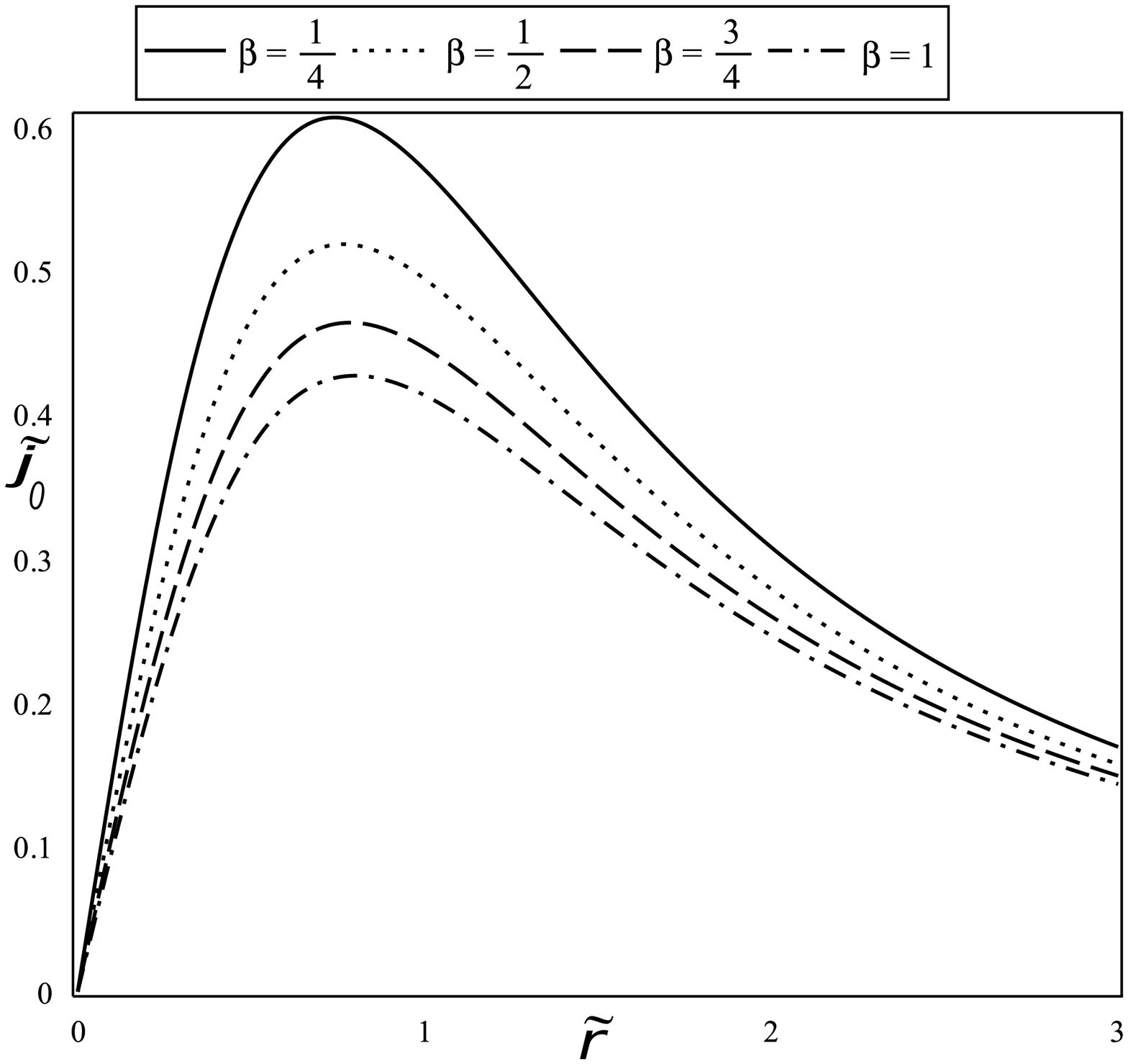}&
\epsfig{width=3in,file=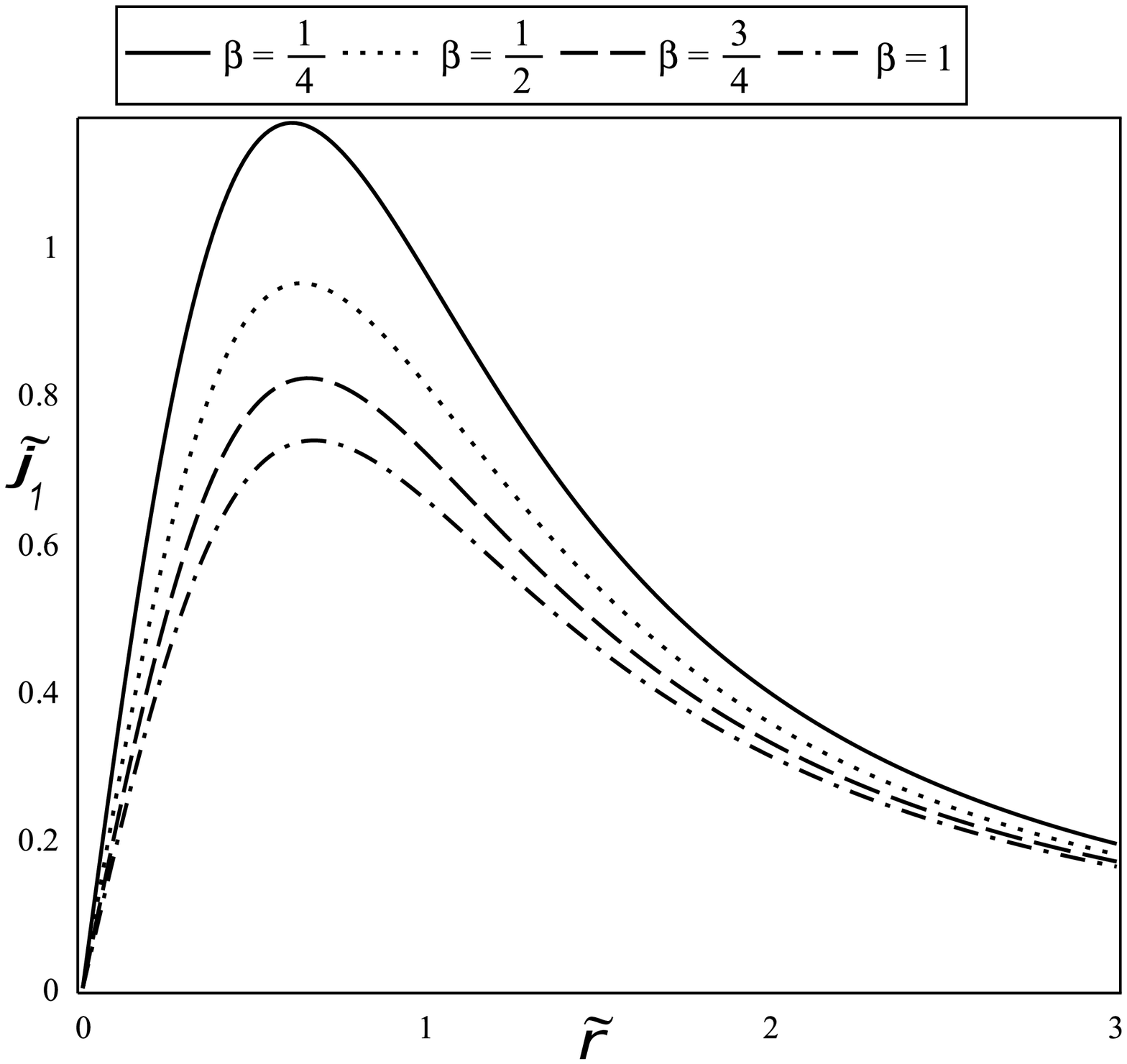}\\
(a) & (b)\\
\end{array}$$
\caption{\label{fig:figure2} Dimensionless  current of magnetization ${\tilde \mu}$  as a function of ${\tilde r}$.
In each case, we plot ${\tilde \mu}_0(\tilde r)$  and ${\tilde \mu}_1(\tilde r)$ for different values of
the parameter $\beta$  with $\tilde{b}_0=1$ and $\tilde{b}_1=0.5.$}
\end{figure}
In Fig. \ref{fig:figure2}, we show the  dimensionless current of magnetization   ${\tilde j}$ as a function of ${\tilde
r}$. In each case, we plot ${\tilde j}_0(\tilde r)$ [Fig. \ref{fig:figure2}(a)]  and ${\tilde j}_1(\tilde r)$ [Fig. \ref{fig:figure2}(b)] for different values of the parameter $\beta$  
with $\tilde{b}_0=1$ and $\tilde{b}_1=0.5$. It can be seen that the  current of magnetization is everywhere positive. It can be observed that 
for all the values of ${\beta}$  
the  current of magnetization is  zero at the  center of the  disk,  increases rapidly as one moves away from the disk center, reaches a maximum and later decreases rapidly. 
It can also be observed that the maximum of the  current of magnetization  increases as the values of the parameter $\beta$ decrease. We also computed the 
functions  $\tilde \mu$ and $\tilde j$ for other values of the parameters and, in all the cases, we found the same behaviour.  We  do not plot  the heat flux, it shows a similar 
behaviour to that of the surface energy density.\\

Before  ending  this  section we  evaluate the  constants  of  motions. From  (\ref{eq:metricpot1}) with $k_3k_4=  - k_5=-1$, $k_8=0$  and   $k_6= -1/k$  we have
         \begin{eqnarray}
                  \phi= \frac{1}{1+ \beta} \ln{\left( \frac{1}{1 - U} \right)}.\label{eq:phi}
          \end{eqnarray}

Then leading term in expansion (\ref{eq:GeneralizedKuzmin}) $U\approx -b_0/R$ determine ADM mass of the spacetime \cite{katz1999disc}       
             \begin{eqnarray}
             M_0= \frac{ b_{_{0}} }{ (1 + \beta)},
               \end{eqnarray}
its total angular momentum
                 \begin{eqnarray}
                           L_{M0} = \frac{1}{2} k_{\omega} {b}_{_{0}}
                           \end{eqnarray}
and  the total  magnetic  dipole moment
                 \begin{eqnarray}
                           L_{B0} = \frac{{b}_{_{0}} }{k}. 
                           \end{eqnarray}
We thus  see that constants $k$  and $k_\omega$  defines the  gyromagnetic ratio $ L_{M0}/ L_{B0} = (k k_{\omega})/2$.

\section{Concluding remarks} \label{sec:conclude}
We  have used  the  formalism presented in   \cite{PhysRevD.87.044010} to obtain  an  exact relativistic  model  describing a  system composed  of  a  thin  disk surrounded by  a 
magnetized halo in a conformastationary space-time background.  The  model  was  obtained by  solving  the Einstein-Maxwell distributional field equations   through the 
introduction of an auxiliary harmonic function that determines the functional dependence of the metric components and the electromagnetic potential under the  assumption that the  
energy-momentum tensor  can be  expressed as  the  sum of  two distributional contributions, one  due to the electromagnetic  part and the  other due to a  material part. As we can 
 see, due to that the spacetime  here considered is non-static (conformastationary) this distributional  approach allows us  a strongly non-linear partial  equation system. We 
have 
considered  for simplicity the astrophysical consistent   case  in that there is  not electric  charge on the halo region. Consequently, it has been obtained  that  the  charge 
density on the  disk region  is zero.

To analyse the physical content of the energy-momentum tensor of the  disks we expressed it in the canonical form  and  we projected it in a comoving frame defined trough of the 
local observers tetrad. This analysis has allowed us  to  give  a  complete  dynamical description  of the  system in terms of two parameters (i.e $\beta$ and $k_\omega$) which  
determine  the  matter  content  of  the  disks. So, in this  paper we  presented  for  first time the complete analysis  of  the most  general energy-momentum tensor of  the  
disks  that can be  obtained  from conformastationary axially symmetric solutions  of  the  Einstein-Maxwell equations.

The expressions obtained here  are the  generalisations of the  corresponding  expressions  for the conformastatic  disks without isotropic pressure, stress tensor or heat flow 
presented in \cite{PhysRevD.87.044010}. Indeed, when the  parameter $\beta$ in the metric  is equal to one  the isotropic pressure and the anisotropic tensor on the  material 
constituting the disks disappear. In a similar way, when  the parameter $k_{\omega}$ is equal to zero the heat flux on the disk vanishes, a feature of the static systems.  
Furthermore, when we take simultaneously $k_{\omega}= 0$ and $\beta =1$, the results   here presented describe the energy-momentum tensor the  disks presented  in  
\cite{PhysRevD.87.044010} for the special case when the electric potential vanishes. Moreover, to illustrate the  application of the formalism  we  have considered  specific 
solutions  in which the gravitational and magnetic potential are completely determined  by  a ``generalization'' of  the  Kuzmin solution of  the Laplace's equation. Accordingly,  
we  have  obtained conformastationary magnetized thin disks of infinite radius, generated  from  a Newtonian gravitational  potential of a static axisymmetric distribution of  
matter. Hence, when a particular  value of  the parameters  $b_0, b_1, \beta, k_\omega$ is taken, the conformastatic  disks without radial pressure presented in 
\cite{PhysRevD.87.044010} are obtained.
  
Since  all  the  relevant quantities show a physically reasonable behaviour,  we  conclude that  the  solution presented here  can  be  useful to  describe the  gravitational  
and  electromagnetic  field  of  a  conformastationary thin  disk surrounded  by  a  halo  in the  presence of  an electromagnetic  field.   In a  subsequent work  we  will  
present a  detailed analysis of  the energy-momentum  tensor  of  the  halo  as  well  as  a thermodynamic analysis of the  disk-halo.

\section*{Acknowledgement}
The author would like to thank the anonymous reviewer for their valuable comments and suggestions to improve the quality of the paper.
The author also wishes to acknowledge  useful discussions  with  C. S. Lopez-Monsalvo and H. Quevedo.
\appendix

\section{The Einstein-Maxwell equations and the thin-disk-halo system}\label{sec:einm}
 Inspired by inverse method techniques, let us assume that the solution has the general form  \cite{synge1960relativity, stephani2003exact}
                      \begin{eqnarray}
                          \mathrm ds^2 = -   f^2( \mathrm dt  +\omega  \mathrm d\varphi)^2 +   \Lambda^4  [\mathrm dr^2 +
                           \mathrm dz^2 + r^2{\mathrm d}\varphi^2  ],  \label{eq:met0}
                             \end{eqnarray}
where we have  introduced  the cylindrical coordinates $x^{\alpha}=(t,r,z,\varphi)$ in which the metric function $f,
\Lambda$ and $\omega$ and  the electromagnetic potential, $A_{\alpha} = (A_t, 0, 0, A_{\varphi})$   depend only on $r$
and $z$ .  Accordingly, the  non-zero components of  the   energy-momentum tensor  of  the  halo,
$M^\pm_{\alpha\beta}=G_{\alpha\beta}^{\pm} - E^\pm_{\alpha\beta}, $  are given by
                          \begin{subequations}\begin{eqnarray}
                               M_{tt}^{\pm}&=&\frac{1}{4r^2\Lambda^8} \{   3f^4\nabla\omega \cdot \nabla\omega -16r^2f^2\Lambda^3
                                                    \nabla^2\Lambda -  2 f^2 \nabla A_{\varphi}\cdot \nabla A_{\varphi} 
                                                    + 4f^2\omega  \nabla A_{t}\cdot \nabla A_{\varphi}\nonumber\\
                                                   &-& 2(r^2\Lambda^4 + f^2\omega^2) \nabla A_{t}\cdot\nabla A_{t} \},\\
                     M_{t\varphi}^{\pm}&=&\frac{1}{4r^2\Lambda^8} \{ -4r^2f^2\Lambda^3\nabla\omega \cdot\nabla\Lambda
                                                    + 3f^4\omega\nabla\omega\cdot\nabla\omega - 16r^2f^2\Lambda^3\omega\nabla^2\Lambda
                                                    + 6r^2\Lambda^4f \nabla\omega \cdot\nabla f \nonumber\\
                                                  &+& 2\Lambda^4r^2f^2 \nabla^2\omega - 4rf^2\Lambda^4\nabla\omega\cdot\nabla r
                                                  -2\omega f^2\nabla A_{\varphi}\cdot \nabla A_{\varphi}
                                                    + 2(\omega\Lambda^4r^2 - f^2\omega^3)\nabla A_{t}\cdot \nabla A_{t} \nonumber\\
                                                  &+& 4(f^2\omega^2 - r^2\Lambda^4 \nabla A_{t}\cdot \nabla A_{\varphi}) \}=M_{\varphi t}^{\pm},\\
                              M_{rr}^{\pm}&=&-\frac{1}{4r^2\Lambda^4f^2} \{ -f^4(\omega_r^2 - \omega_z^2)- 4\Lambda^4r^2f\nabla^2f
                                                    + 4r^2f\Lambda^4f_{rr} - 8r^2f^2\Lambda^3\nabla^2\Lambda  + 8r^2f^2\Lambda^3\Lambda_{rr} \nonumber\\
                                                 & - & 16r^2\Lambda^3ff_{,r}\Lambda_{,r} + 8f^2\Lambda^2r^2\Lambda_{,z}^2
                                                    - 16f^2r^2\Lambda^2\Lambda_{,r}^2 + 2f^2 ( A_{\varphi,r}^2 -  A_{\varphi,z}^2)
                                                    - 2(\Lambda^4r^2 - f^2\omega^2)(A_{t,r}^2 - A_{t,z}^2) \nonumber\\
                                                 & -& 4\omega f^2(A_{t,r}A_{\varphi,r} - A_{t,z}A_{\varphi,z}  )  \},\\
                            M_{rz}^{\pm}&=&\frac{1}{2r^2\Lambda^4f^2} \{ 4r^2f\Lambda^3(\Lambda_{,r}f_{,z} + \Lambda_{,z}f_{,r}) 
                                                  +  12r^2f^2\Lambda^2\Lambda_{,r}\Lambda_{,z}  + f^4\omega_{,r}\omega_{,z} - 4r^2f^2\Lambda^3\Lambda_{,rz}\nonumber
                                             \\& - & 2r^2\Lambda^4ff_{,rz} - 2(f^2\omega^2 - \Lambda^4r^2) A_{t,r}A_{t,z}
                                                 + 2f^2\omega(A_{\varphi,r}A_{t,z} +A_{t,r}A_{\varphi,z}) -2f^2A_{\varphi,r}A_{\varphi,z} \},\\
                         M_{zz}^{\pm}&=&\frac{1}{4r^2\Lambda^4f^2} \{-f^4 (\omega_{,r}^2   - \omega_{,z}^2  ) + 4r^2\Lambda^4f\nabla^2f
                                                 - 4r^2\Lambda^4ff_{,zz} + 8r^2f^2\Lambda^3\nabla^2\Lambda -8r^2f^2\Lambda^3\Lambda_{,zz}  \nonumber\\
                                              &-& 8r^2f^2\Lambda^2\Lambda_{,r}^2   + 16r^2\Lambda^3ff_{,z}\Lambda_{,z}
                                               + 16r^2f^2\Lambda^2\Lambda_{,z}^2 + 2f^2(A_{\varphi,r}^2  - A_{\varphi,z}^2 )
                                               - 4\omega f^2 (A_{t,r}A_{\varphi,r} - A_{t,z}A_{\varphi,z}    ) \nonumber\\
                                            & - & 2(\Lambda^4r^2 - f^2\omega^2)(A_{t,r}^2  - A_{t,z}^2) \},\\
      M_{\varphi \varphi}^{\pm}&=&\frac{1}{4r^2\Lambda^8f^2} \{4r^4f\Lambda^8\nabla^2f 
                                                 - 4r^3\Lambda^8f\nabla  f\cdot\nabla r + f^4(\Lambda^4r^2  +3\omega^2f^2)\nabla \omega\cdot\nabla\omega
                                                 -  16r^2f^4\omega^2\Lambda^3\nabla^2\Lambda  \nonumber\\
                                              & + & 4r^2\Lambda^4f^4\omega\nabla^2\omega - 8r^4f^4\omega\nabla\omega\cdot\nabla r  
                                              + 12\Lambda^4r^2\omega f^3\nabla\omega\cdot\nabla f
                                               - 8r^2\omega f^4\Lambda^3\nabla \omega\cdot\nabla\Lambda \nonumber\\
                                              &+& 8f^2\Lambda^7r^4\nabla^2 \Lambda - 8f^2\Lambda^7r^3\nabla \Lambda\cdot\nabla r
                                               - 8r^4 f^2 \Lambda ^6 \nabla \Lambda\cdot\nabla \Lambda
                                              - 2(f^2\Lambda^4r^2 + f^4\omega^2)\nabla A_{\varphi}\cdot\nabla A_{\varphi}\nonumber\\
                                            &-& 4(f^2\omega\Lambda^4r^2 - \omega^3f^4)\nabla A_{\varphi}\cdot\nabla A{t}
                                              -2(\Lambda^8r^4 -2\Lambda^4r^2f^2\omega^2 + f^4\omega^4)\nabla A_{t}\cdot\nabla A_{t} \} .
                         \end{eqnarray}\label{eq:haloEMT}\end{subequations}
 Furthermore,  the  electric current density  of  the  halo  reads
                    \begin{subequations}\begin{eqnarray}
                       J^{t}_{\pm}&=&\frac{1}{\Lambda^6f} \nabla \cdot \{\Lambda^2 f^{-1}\nabla A_{t} + r^{-2}\omega f\Lambda^{-2} (\nabla A_{\varphi} 
                       - \omega\nabla A_{t})   \},\\
              J^{\varphi}_{\pm}&=&\frac{1}{\Lambda^6f} \nabla \cdot \{ r^{-2}f\Lambda^{-2} (\nabla A_{\varphi} - \omega \nabla A_t  ) \}.
           \end{eqnarray}\label{eq:haloCURRENT}\end{subequations}
The  discontinuity in  the  $z$-direction of $Q_{\alpha\beta}$  and  ${\cal I}^{\alpha}$ defines,  respectively,  the surface
energy-momentum tensor (SEMT)  and the surface electric  current density (SECD)  of  the  disk $S_{\alpha\beta}$, more precisely
                    \begin{subequations}\begin{eqnarray}
                        S_{\alpha\beta} &\equiv&   \int Q_{\alpha\beta}  \delta(z)  ds_n  = \sqrt{g_{zz}}  Q_{\alpha\beta} ,\\
                              {\cal J}^{\alpha} &\equiv&   \int {\cal I}^{\alpha}  \delta(z)  ds_n  = \sqrt{g_{zz}}  {\cal I}^{\alpha}
                           \end{eqnarray}\end{subequations}
where $ds_n = \sqrt{g_{zz}} \ dz$ is the ``physical measure'' of length in the direction normal to the $z = 0$ surface.
Accordingly, the  non-zero components of  the SEMT for  the  line  element (\ref{eq:met0}) are  given by
         \begin{subequations}\begin{eqnarray}
              S^t_{\; t}   &=&\frac{1}{r^2\Lambda^6}\{ -f^2\omega\omega_{,z} + 8r^2\Lambda^3\Lambda_{,z} \}\\
              S^t_{\; \varphi}   &=&-\frac{1}{r^2 \Lambda^6f} \{ -4r^2\omega f\Lambda^3  \Lambda_{,z}
                  + \omega^2 f^3 \omega_{,z} + 2 r^2\Lambda^4\omega f_{,z}    + r^2 f \Lambda^4 \omega_{,z}  \},\\
              S^r_{\; r}   &=&\frac{2}{\Lambda^3f}\{  2f\Lambda_{,z} + \Lambda f_{,z}  \},\\
              S^{\varphi}_{\;\; t} &=&\frac{f^2}{r^2\Lambda^6}  \omega_{,z}\\
              S^{\varphi}_{\; \varphi}   &=& \frac{1}{r^2\Lambda^6 f}\{ 2 r^2\Lambda^4 f_{,z}
              +  4r^2\Lambda^3  f \Lambda_{,z}+ \omega f^3 \omega_{,z}  \},
                  \end{eqnarray}\label{eq:SEMT}\end{subequations}
whereas the  non-zero  components  of  the (SECD) are
              \begin{subequations}\begin{eqnarray}
                  {\cal J}^{t}&=&\frac{1}{r^2\Lambda^6 f^2} \{ \omega f^2 \left[  A_{\varphi,z}\right]  +  (r^2\Lambda^4   - f^2 \omega^2)   \left[  A_{t ,z}\right]   \},\\
                  {\cal J}^{\varphi}&=&\frac{1}{r^2\Lambda^6 } \{ \omega \left[  A_{t ,z}\right] - \left[  A_{\varphi,z}\right]   \},
                  \end{eqnarray}\label{eq:diskCURRENT}\end{subequations}
where all  the  quantities  are evaluated on the  surface of  the  disk.

\section{The local  observers}
We  write the  metric (\ref{eq:met1}) in the form
      \begin{eqnarray}
          \mathrm ds^2 = -   F( \mathrm dt  + \omega  \mathrm d\varphi)^2 +   F^{-\beta}  [\mathrm dr^2 + \mathrm  dz^2 + r^2{\mathrm d}\varphi^2  ], \label{eq:met2}
             \end{eqnarray}
where we have rewritten  $F=e^{2\phi}$.  The tetrad of the  local observers $h^{(\alpha)}_{\;\;\mu}$, in which the  metric  has  locally the  form  of Minkowskian metric
       \begin{eqnarray}
           ds^2=\eta_{(\mu)(\nu)}\mathbf{h}^{( \mu)}\otimes\mathbf{h}^{ (\nu)},
          \end{eqnarray}
is  given  by
          \begin{subequations}\begin{eqnarray}
                {h}^{(t)}_{\quad \alpha}&=&F^{1/2}\{ 1,0,0, \omega \},\\
                {h}^{(r)}_{\quad\alpha}&=&F^{-\beta/2}\{ 0,1,0, 0 \},\\
                {h}^{(z)}_{\quad\alpha}&=&F^{-\beta/2}\{ 0,0,1,0 \},\\
                {h}^{(\varphi)}_{\quad\alpha} &=&F^{-\beta/2}\{ 0,0,0,r \}.\\
                \end{eqnarray}\label{eq:localobservator}\end{subequations}
The dual  tetrad reads
               \begin{subequations}\begin{eqnarray}
               {h}_{(t)}^{\quad \alpha}&=&F^{-1/2}\{ 1,0,0, 0 \},\\
               {h}_{(r)}^{\quad\alpha}&=&F^{\beta/2}\{ 0,1,0, 0 \},\\
               {h}_{(z)}^{\quad\alpha}&=&F^{\beta/2}\{ 0,0,1,0 \},\\
               {h}_{(\varphi)}^{\quad\alpha} &=&\frac{F^{\beta/2}}{r}\{ -\omega,0,0,1 \}.\\
                     \end{eqnarray}\label{eq:duallocalobservator}\end{subequations}

The  circular velocity of  the system disk-halo can be  modelled  by a fluid space-time whose circular velocity
$V^{\alpha}$ can be written in terms of two Killing  vectors $t^{\alpha}$ and $\varphi^{\alpha}$,
            \begin{eqnarray}
                V^{\alpha}=V^t(t^{\alpha} + \Omega\varphi^{\alpha})\label{eq:velocity},
                \end{eqnarray}
where
            \begin{eqnarray}
               \Omega \equiv\frac{u^{\varphi}}{u^t}=\frac{d\varphi}{dt}
                  \end{eqnarray}
is  the  angular velocity  of  the  fluid  as  seen  by  an  observer at rest at infinity. The  velocity satisfy the  normalization $V_{\alpha}V^{\alpha}=-1$.
Accordingly for  the metric (\ref{eq:met2} ) we  have
             \begin{eqnarray}
                (V^{t})^2=\frac{1}{-t^{\alpha}t_{\alpha} - 2\Omega t^{\alpha}{\varphi}_{\alpha} - \Omega{\varphi}^{\alpha}{\varphi}_{\alpha}},\label{eq:cuadrivelocity}
               \end{eqnarray}
with
             \begin{subequations}\begin{eqnarray}
                t^{\alpha}t_{\alpha}&=&g_{tt}=- F\\ 
                t^{\alpha}{\varphi}_{\alpha}&=&g_{t\varphi}=-\omega F\\
                {\varphi}^{\alpha}{\varphi}_{\alpha}&=&g_{\varphi\varphi}=r^2F^{-\beta}(1 - F^{1+\beta}\frac{\omega^2}{r^2}).
                     \end{eqnarray}\end{subequations}
Consequently, we  write  the  velocity  as
            \begin{eqnarray} 
               V^t=\frac{1}{F^{1/2}(1 +\omega\Omega)\sqrt{1 - V_{_{LOC}}^2}},
                 \end{eqnarray}
where
            \begin{eqnarray}
                 V_{_{LOC}}\equiv \frac{r\Omega}{F^{{(1+\beta)}/{2}}(1 +\omega\Omega)},\label{eq:localvelocity}
                 \end{eqnarray}
 is the  velocity as measured  by  the local observers.

       
       %
       \end{document}